\documentclass[12pt]{article}
\usepackage{graphicx}%
\usepackage{amsmath}
\begin{document}
%
\begin{center}
{\bfseries STUDY OF TWO-PHOTON PRODUCTION PROCESS IN PROTON-PROTON
COLLISIONS BELOW THE PION PRODUCTION THRESHOLD}
\vskip 5mm
{\large \bf A.S.Khrykin}\\[0.5 cm]
{ \small \it Joint Institute for Nuclear Research, Dubna, 141980
Russia\\
Email: Khrykin@nusun.jinr.dubna.su}
\end{center}
\begin{abstract}
The energy spectrum for high
energy $\gamma$-rays ($E_\gamma \geq 10$ MeV) from the process $pp
\to \gamma \gamma X$ emitted at $90^0$ in the laboratory frame
has been measured at an energy below the pion
production threshold, namely, at 216 MeV. The resulting photon energy
spectrum extracted from $\gamma - \gamma$ coincidence events
consists of a narrow peak at a photon energy of
about 24 MeV and a relatively broad peak in the energy range of
(50 - 70) MeV. The statistical significance for both the
narrow and broad peaks exceeds 5$\sigma$. This behavior of
the photon energy
spectrum is interpreted as a signature of the exotic dibaryon
resonance $d^\star_1$ which is assumed to be formed in the radiative
process $pp \to \gamma \ d^\star_1$ followed by its electromagnetic decay
via the
$d^\star_1 \to pp \gamma$ mode.
The experimental spectrum is compared with those obtained by means
of Monte Carlo simulations.
\end{abstract}
%
%
\section{Introduction}
Direct-production of two hard photons in nucleon-nucleon
collisions ($NN\gamma\gamma$) at intermediate energies, unlike
production of single photons ($NN$ bremsstrahlung), belongs to those
fundamental processes which are still poorly explored both
theoretically and experimentally.
However, a study of this process
can provide us with new important information on the underlying mechanisms
of the $NN$ interaction that is complementary to that obtained from
investigations of other processes. In particular, the
$NN\gamma\gamma$ process can be used as a sensitive probe for
experimental verification of the possible existence of
$NN$-decoupled nonstrange dibaryon resonances. These are two-baryon
states $^2B$ with zero strangeness and exotic quantum numbers,
for which the strong decay
$^2B \to NN$ is either strictly forbidden by the Pauli
principle (for the states with the isospin $I=1(0)$ and with an even
sum $L+S+I$, where $L$ is the orbital momentum and $S$ is the spin),
or is strongly
suppressed by the isospin selection rules (for the states with
$I=2$)\cite{GerKh93,EGK95}. Such dibaryon
states cannot be simple bound systems
of two color-singlet nucleons, and
a proof of their existence would have consequences of fundamental
significance for the theory of strong interactions.
The $NN$-decoupled dibaryon states are predicted in a series
of $QCD$-inspired models\cite{QCDinm,Kondr,Kopel}.
Among the predicted dibaryons, there are those whith masses both
below and above the pion production ($\pi NN$) threshold.
The $NN$-decoupled dibaryons with masses below the $\pi NN$ threshold
are of special interest,
since they may decay mainly into the $NN\gamma$ state, and
consequently, their widths should be very narrow($\le 1 keV$).
Narrow dibaryon states have been searched for
in a number of experiments\cite{Tat00}, but none has provided any undoubted evidence
for their existence.
Most of the past searches, however, were
limited to dibaryons coupled to the $NN$ channel, and,
to our knowledge, only a few ones were  dedicated to the
$NN$-decoupled dibaryons\cite{Clement,Konob}.
At the same time, since the latter do not couple to the
$NN$-channel, then, depending on their production and decay modes,
they may have escaped detection up to now and may naturally appear
in dedicated experiments.\\
If the $NN$-decoupled dibaryons
exist in nature, then the $NN\gamma\gamma$ process may proceed, at
least partly, through the mechanism that directly involves the
radiative excitation $NN \to \gamma \ ^2B$ and decay $^2B \to
\gamma NN$ modes of these states.
In $NN$ collisions at energies below the $\pi NN$
threshold, these production and decay modes of
the $NN$-decoupled dibaryon resonances with masses
$M_R \leq 2m_N + m_\pi$ would be unique or dominant.
The simplest and clear way of revealing them
is to measure the photon energy spectrum of the
reaction $NN\gamma\gamma$. The presence of an $NN$-decoupled dibaryon
resonance would reveal itself in this energy spectrum as a narrow line associated with
the formation of this resonance and a relatively
broad peak originating from its three-particle decay.
In the center-of-mass system, the position of the narrow line($E_R$)
is determined by the energy of colliding nucleons ($W=\sqrt{s}$)
and the mass of this dibaryon resonance as $E_R=(W^2-M_R^2)/2W$.
An essential feature of the two-photon production in $NN$
collisions at an energy below the $\pi NN$ threshold is that,
apart from the resonant mechanism in question, there should only
be one more source of photon pairs.
This is the double $NN$-bremsstrahlung reaction.
But this reaction is expected to play a minor role.
Indeed, it involves two electromagnetic vertices, so that one may
expect that the $NN\gamma\gamma$-to-$NN\gamma$ cross section ratio
should be of the order the fine structure constant $\alpha$.
However, the cross section for $NN\gamma$ is already
small (for example, the total $pp$-bremsstrahlung cross section at energies
of interest is a few $\mu b$).\\
The process $pp \to pp \gamma
\gamma$ at an energy below the $\pi NN$ threshold provides
an unique possibility of searching for the $NN$-decoupled dibaryon
resonances in the mass region $M_R \leq 2m_p+m_\pi$ with
quantum numbers $I(J^P)=1(1^+,3^+,etc.$) or those with $I=2$
($J$ is the total spin, and $P$ is the parity of a dibaryon).
The preliminary experimental studies of the
reaction $pp \to \gamma\gamma X$ at an incident proton
energy of about 200 MeV\cite{Panic96,Menu97} showed that the photon energy
spectrum of this reaction had a peculiar structure ranging from
about 20 MeV to about 60 MeV. This structure was interpreted as
an indication of the possible existence of an $NN$-decoupled dibaryon
resonance(later called $d^\star_1$) that is produced
in the process $pp \to \gamma \ d^\star_1$ and subsequently decays via
the $d^\star_1 \to pp \gamma$ channel.
Unfortunately, a relatively
coarse energy resolution and low statistics did not allow
us to distinguish the narrow $\gamma$-peak associated with
the $d^\star_1$ production from the broad $\gamma$-peak due to the decay
of this resonance and, hence, to determine the resonance mass exactly.
To get a rough estimate of the $d^\star_1$ mass, we admitted that
the expected narrow $\gamma$-peak is positioned at the center of
the observed structure and thus obtained $M_R \sim 1920$ MeV.
However, if the uncertainty in the position of the narrow $\gamma$-peak
in the structure is taken into account,
the actual $d^\star_1$ mass might be considerably
different from such a crude estimate.
The possible existence of this dibaryon resonance was also probed in
the proton-proton bremsstrahlung data taken by
the WASA/Promice Collaboration at the CELSIUS accelerator\cite{WASA}
for proton energies of 200 and 310 MeV.
However, the trigger used in those experimental studies to select events
was designed and optimized for investigating the usual $pp \to pp\gamma$
bremsstrahlung. The selected events were those
corresponding to two simultaneously detected protons emerging in the forward
direction at angles between $4^0$ and $20^0$ with respect to the
beam axis, each of which had a kinetic energy exceeding the present
threshold (40 MeV). The analysis of those data resulted only in upper
limits on the dibaryon production cross section
in the mass range from 1900 to 1960 MeV.
Therefore, to clarify the situation with the dibaryon
resonance $d^\star_1$, we have decided to measure the energy spectrum
of the $pp \to pp\gamma\gamma$ reaction more carefully.
\section{Experimental apparatus and event selection}
The experiment was performed using the variable energy proton
beam from the phasotron at the Joint Institute for Nuclear
Research(JINR). The schematic layout of the experimental setup is
shown in Fig. 1.
\begin{figure}[t]
\begin{center}
\includegraphics[width=80 mm]{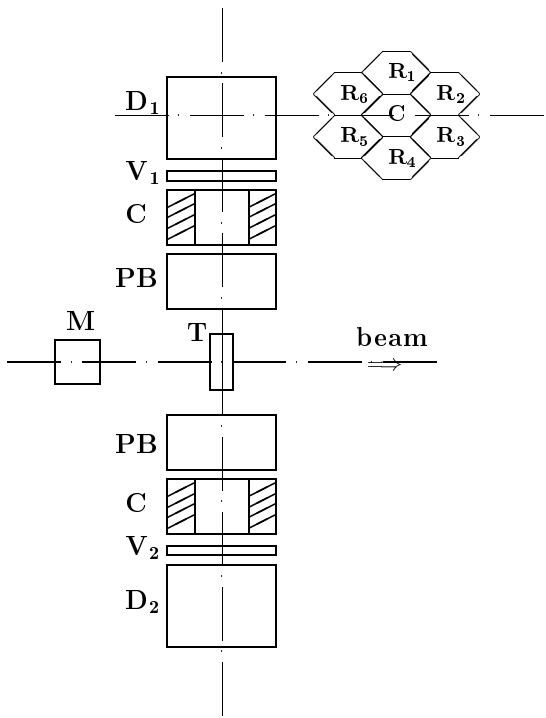}%
\caption{Schematic layout of the experimental setup}
\end{center}
\end{figure}
 The pulsed proton beam (bursts of about
7 ns FWHM separated by 70 ns) with an energy of about 216 MeV and
an energy spread of about 1.5\%
bombarded a liquid hydrogen target ($T$).
 The average beam intensity was on the order of $3.6 \cdot 10^8$ protons/s
and was monitored by an ionization chamber ($M$).
The liquid hydrogen target was a cylindrical cryogenic vessel with
a length of about 5 cm ($\sim 0.35 g/cm^2$) and a diameter of 4.5,
cm which had thin
entrance and exit windows made from 100 $\mu m$ thick mylar foil.
Both $\gamma$-quanta of the reaction $pp\gamma\gamma$ were detected by
two $\gamma$-ray detectors($D_1$ and $D_2$) placed
in a horizontal plane, symmetrically on either side of the beam
at a laboratory angle of $90^0$ with respect
to the beam direction. The $\gamma$-detector
$D_1$ was an array of seven individual detector modules (see Fig. 1):
the central detector module ($C$) and six detector modules ($R_1,
R_2,..,R_6$) surrounding the central detector module. Each
detector module was a 15 cm thick $CsI(Tl)$ crystal having a hexagonal
cross section with an outer diameter of 10 cm. The detector modules
$R_1,R_2,..,R_6$ (ring) were designed to measure the energy of
secondary particles and photons leaving the central detector module $C$.
Thus, the energy of each photon detected by the detector $D_1$ was
determined by summing up the energy deposited in the central
detector module and the energies deposited in the detector modules of
the ring.
Besides, the ring was used as an active shielding from cosmic ray
muons. The energy
resolution of a single $CsI(Tl)$ crystal was measured to be
about 13\% for $E_\gamma = 15.1$ MeV.
 The second $\gamma$-ray detector $D_2$ was a cylindrical
$NaI(Tl)$ crystal 15 cm in diameter and 10 cm in length.
Compared to the detector $D_1$, it had a poorer energy
resolution for $\gamma$-rays with energies $E_\gamma > 10$ MeV.
The detector $D_2$ was mainly designed to select $\gamma - \gamma$
coincidence events.
To reject events induced by charged particles, plastic scintillators ($V_1$
and $V_2$) were put in front of each $\gamma$-detector. Both
$\gamma$-ray detectors were placed in lead houses with walls 10 cm
thick which were in turn surrounded by a 5 cm borated
parafin shield. The solid angles covered by
the detectors $D_1$ and $D_2$ were 43 msr and 76 msr, respectively.
To attenuate prompt particles coming from the target, 15 cm
polyethylene bars (PB) were placed in between the target and each
$\gamma$-detector.\\
The electronics associated with the
$\gamma$-detectors and the plastic scintillators provided amplitude
and time signals suitable for digitizing by an $ADC$ and a $TDC$.
The data acquisition system was triggered by the time signal from
the detector module $C$ which started
all the $TDCs$ and  opened the $ADC$ gates.
In parallel, the time signals from the detector module $C$ and the
$\gamma$-detector $D_2$ were sent to a coincidence circuit to select
those events
which had signals from both the $\gamma$-detectors.
Events satisfying this criterion were recorded in the event-by-event mode
on the hard disk of the computer.
A further selection of $\gamma - \gamma$ coincidence
candidate events was done during off-line data processing.

The off-line selection procedure for events associated with
the process $pp \to \gamma \gamma X$ included the following main
operations:
rejection of events induced by charge particles coming from the
target,
rejection of events induced by cosmic ray muons,
selection of events which are caused by proton beam bursts and
selection of events associated with $\gamma - \gamma$ coincidences.
\section{Results and discussion}
The measurements were performed both with
the target filled with liquid hydrogen and with
the empty one. Data for the full and empty targets were taken
in two successive runs for $\sim$31 h and $\sim$21 h, respectively.
The integrated luminosity of about $8.5\  pb^{-1}$ was
accumulated for the measurement with the full target.
The resulting number of $\gamma - \gamma$ coincidence events
associated with the $pp \to \gamma \gamma X$ process
as a function of the photon energy measured by the detector $D_1$
is presented in Fig. 2. As can be seen, the energy spectrum of
the reaction $pp \to \gamma \gamma X$
consists of a narrow peak at a photon energy of about 24
MeV and a relatively broad peak in the energy range from about 50 to
about 70 MeV. It was found that the statistical significance for the sharp peak as well
as for the broad
one exceeds 5$\sigma$. The width (FWHM) of the narrow peak
was found to be about 8 MeV. This width is comparable with that
of the energy resolution of the experimental setup.
The observed behavior of the photon energy
spectrum agrees with a characteristic signature of the
sought dibaryon resonance $d^\star_1$ that is formed and decays
in the radiative process $pp \to \gamma \ d^\star_1 \to pp \gamma\gamma$.
In that case the narrow
peak should be attributed to the formation of
this dibaryon, while the broad peak should be assigned
to its three-particle decay. Using
the value for the energy of the narrow peak $E_R \sim 24$ MeV, we obtained
the $d^\star_1$ mass  $M_R \sim 1956$ MeV.

Having assumed that the $pp \to \gamma \ d^\star_1 \to pp
\gamma\gamma$ process with the $d^\star_1$ mass of 1956 MeV is the only
mechanism of the reaction $pp \to pp \gamma\gamma$, we calculated
the photon energy spectra of this reaction for a proton energy of
216 MeV. It was also assumed that the
radiative decay of the $d^\star_1$ is a dipole $E1(M1)$ transition
from the two-baryon resonance state to a $pp$-state in the continuum.
The calculations were carried out with the help of Monte Carlo
simulations which included the geometry and the energy resolution of
the actual experimental setup. The photon energy spectra were
calculated for two different scenarios of the $d^\star_1$ decay.
The difference between them was that one of these scenarios took into
account the final state interaction ($FSI$) of two outgoing
protons whereas in the other that interaction was switched off.
Each of the
scenarios imposed some restrictions on possible quantum numbers of
the dibaryon state in question. The scenario including the $FSI$
implies that the final $pp$-system is in the singlet $^1S_0$
state and consequently it should take place, in particular, for the isovector
$1^+$ dibaryon state (the simplest exotic quantum numbers), namely,
$1^+ \stackrel{M1}{\longrightarrow} 0^+$. At the same time, the scenario
in which the $FSI$ is switched off, is most likely to occur for
the isotensor $0^+$ dibaryon state.
The spectra calculated for these two decay scenarios
 and normalized to the total number of $\gamma - \gamma$
events observed in the present experiment are shown in Fig. 2.

\begin{figure}[t]
\begin{center}
\includegraphics[width=80 mm]{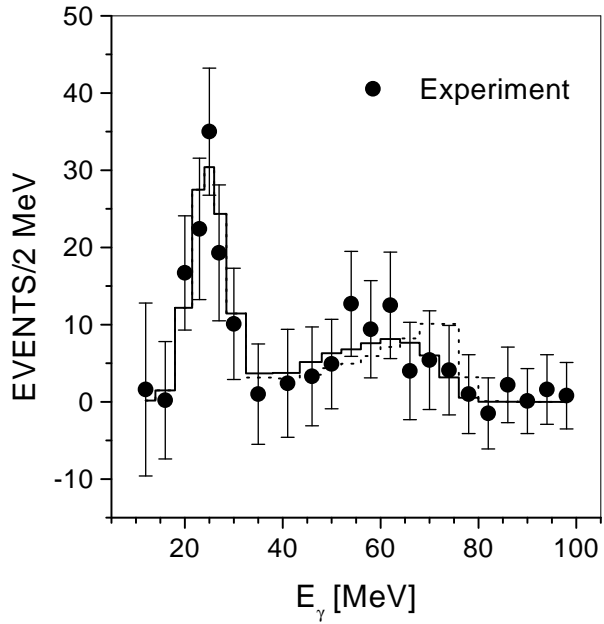}
\caption{\small{Experimentally observed
energy spectrum for photons from the $pp\gamma\gamma$
process and energy spectra for photons from the process
$pp \to \gamma d^\star_1 \to \gamma\gamma pp$
calculated with the help of Monte Carlo simulations for two $d^\star_1$
decay scenarios: without the FSI (solid line)
and with the FSI (dashed line).}}
\end{center}
\end{figure}
Comparison of these spectra with the experimental spectrum indicates
that both the calculated spectra are in reasonable agreement with
the experimental one within experimental uncertainties. However,
one can see that the energy spectrum for the scenario that
can be realized for the isotensor dibaryon($I=2$) is more consistent
with the experimental spectrum. At the same time, the statistics of the
experiment is insufficient to draw any firm conclusions in favor
of one of these scenarios, that is, in favor of either $I=1$ or
$I=2$ for the isospin of $d^\star_1$. In this connection, we
would like to emphasize that the experimental data on the
process $\pi^- d \to \gamma\gamma X$ can shed some light on the
properties of the dibaryon state we have observed. The point is
that the $d^\star_1$ dibaryon can in principle be excited in
this process via the mode $\pi^- d \to \gamma d^\star_1 \to \gamma\gamma nn$.
However, excitation of the dibaryon $d^{\ast}_{1}$ with isospin
$I=1$ is allowed by the isospin selection rules, whereas excitation of
the isotensor $(I=2)$ resonance is expected to be strongly
suppressed\cite{Ge97}. Therefore, the preliminary results of the study of the
reaction $\pi^- d \to \gamma\gamma X$ carried out by the RMC
Collaboration at TRIUMF\cite{Triumf} indicating no
enhancement of the two-photon yield in the process $(\pi^{-} d)_{atom}
\to \gamma \gamma nn$ compared to the usual one-nucleon
mechanism $\pi^{-}p \to \gamma\gamma n$ on the proton bound in the deuteron
may imply that the $d^{\ast}_{1}$ dibaryon has isospin 2.
\section{Conclusion}
 The $\gamma$-ray energy spectrum for the $pp \to \gamma \gamma X$ reaction
at a proton energy below the pion production threshold has been measured
for the first time.
The spectrum measured at an energy of about 216 MeV
for coincident photons emitted at an angle of $90^0$ in the laboratory frame
clearly evidences the existence of the $NN$-decoupled dibaryon
resonance $d^\star_1$ with a mass of about 1956 MeV that is
formed and decays in the process $pp \to \gamma \ d^\star_1 \to pp \gamma\gamma$.
The data we have obtained, however, are still incomplete, and
additional careful studies of the reaction $pp \to pp \gamma \gamma$
are needed to get proper parameters (mass, width, spin, etc.)
of the observed dibaryon state.

\end{document}